# Momentary increase of reactive surface area in dissolving carbonate

Y. Yang,[*,1] M. Rogowska,[1] Y. Zheng,[2] S. Bruns,[1] C. Gundlach,[2] S. L. S. Stipp[1] and H. O. Sørensen[1]

[1]Nano-Science Center, Department of Chemistry, University of Copenhagen, DK-2100 Copenhagen, Denmark

[2]Department of Physics, Technical University of Denmark, DK-2500 Lyngby, Denmark

[*]Corresponding author, +45 9161 4575, yiyang@nano.ku.dk

## TOC Art

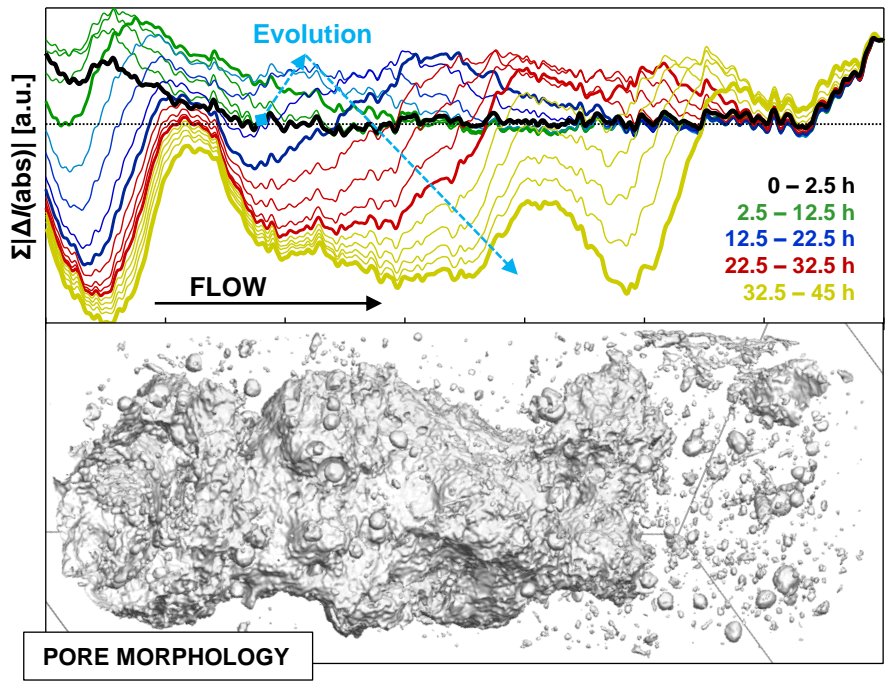

## ABSTRACT

The quantification of surface area between mineral and reactive fluid is essential in environmental applications of reactive transport modelling. This quantity evolves with microstructures and is difficult to predict because the mechanisms for the generation and destruction of reactive surface remain elusive. The challenge of accounting for the inherent heterogeneities of natural porous media in numerical simulation further complicates the problem. Here we first show a direct observation of reactive surface generation in chalk under circumneutral to alkaline pH using *in situ* X-ray microtomography. The momentary increase of reactive surface area cannot be explained by a change in fluid accessibility or by surface roughening stemming from mineralogical heterogeneity. We then combine greyscale nanotomography data with numerical simulations to show that similar temporal behaviour can be observed over a wide range of pH as porous media dissolve in imposed flow field. We attribute the observation to the coupling between fluid flow and mineral dissolution and argue that the extent of surface generation is strongly correlated with the advective penetration depth of reactants. To conclude, we demonstrate the applicability of using a macroscopic Damköhler number as an indicator for the phenomenon and discuss its environmental significance beyond geologic carbon storage.

**INTRODUCTION**

A great challenge in many environmental applications of reactive transport modelling is to account for the microstructural evolution of natural porous media.[1, 2] This evolution depends on both the active physical and chemical processes[3-5] and the characteristic heterogeneities of the media.[6-9] Structural changes influence the variations of petrophysical parameters over time.[10-14] As model inputs, these parameters determine the outcomes of numerical simulations and thus affect the accuracy of predictions.[15, 16]

Reactive surface area (RSA) is among the most important petrophysical parameters that define the outcome of water-rock interactions.[17-20] Conventionally in reactive transport modelling, RSA can be assigned a constant value based on *ex situ* measurements, or be allowed to drift as a function of porosity given by empirical correlations.[18, 21] In recent years, there have been an increasing number of investigations on the temporal behavior of petrophysical parameters, including RSA, in natural materials that are interacting with reactive fluids.[22-26] Both increase and decrease of RSA overtime have been reported and a number of pore scale interpretations are proposed. An important conceptual model is the "sugar lump" model by Noiriel *et al*.[18] The model describes a momentary increase in reactive surface as a result of changing fluid accessibility. In this model, a significant portion of mineral surface does not have access to reactive fluid until the geometric barrier (the "sugar coating") dissolves. The *sugar lump* model explains why RSA may increase whereas the geometric surface area (GSA) remains largely constant or decreases. Similarly, a decrease of RSA can be expected if certain pore throat is clogged by drifting particles in the flow field. Meanwhile, analysis of Budek *et al.* shows that the expansion of cylindrical channels in a pore network can also cause an increase of RSA before solid depletion.[27] This result is important in that it shows the overall GSA can increase within a confined volume, despite the competition and merging of simultaneously developing pores. An increase of RSA in this case indicates the generation of new surfaces and is therefore fundamentally different from the *sugar lump* interpretation. It is, however, not clearly yet if this analytical result, derived from idealized geometry, can be extended to complex natural microstructures.

Yang *et al.* reported a receding dissolution front in an imposed flow field under elevated $CO_2$ pressure.[28] The authors used *in situ* X-ray microtomography (μCT) to monitor the progress of structure disintegration and attributed the recession to a momentary increase of reactive surface. The authors noted that this increase of RSA could not be explained by the *sugar lump* model because the μCT captured the geometric surfaces even when they had not fluid access. The phenomenon was also distinct from the one reported by Budek *et al.* because the fluid-solid interface was defined implicitly on a greyscale basis and the displacement of solid-fluid interface leading to the expansion of individual pores was not tracked.[29] Although the study aims at geologic carbon storage (GCS)-relevant scenarios where the flowing fluid, acidified by $CO_2$, dissolves rock and drives the evolution of microstructure, the authors conclude that their observation stems from the coupling between fluid flow and mineral dissolution. [30, 31] This coupling tends to enhance local porosity contrasts before solid depletion and therefore generates new surface for water-rock interactions. An important implication is that this phenomenon does not rely on the presence of acidic fluid and may be applicable to a very wide range of environmental applications where pH is circumneutral to alkaline. The only prerequisite is the presence of chemical processes leading to a reduction of solid volume with fluid flow.[32] Example environmental scenarios include the *in situ* remediation of organic contaminants using green rust[33, 34] and the safe disposal of nuclear wastes.[35]

In this paper we first use *in situ* μCT to show that the increase of reactive surface can indeed be observed under a circumneutral condition, where the fluid pH increased from ~7 to ~10. We then use a reactor network model to demonstrate the importance of the advective penetration depth of reactant on the extent of RSA increase. The model allows us to keep the Péclet number and the initial microstructure constant while investigate the effects of flowrate and solution composition. Finally, we propose a definition of macroscopic Damköhler number where the pseudo first order rate constant reflects the advective penetration depth of reactive fluid and discuss its generality and environmental significance.

**MATERIALS AND METHODS**

**Experimental**

The dissolution process was monitored by continuous X-ray imaging using a benchtop microtomography system (Versa XRM410, Carl Zeiss X-ray Microscopy, Pleasanton, CA, USA). A cylindrical chalk sample (ø×L = 0.9×2 mm) was machined from an outcrop of Maastrichtian chalk, collected near Aalborg, Denmark (Rørdal Quarry). The sample was loaded in a mini Hassler core holder.[36] Ultrapure water was forced through the sample with a constant flowrate of 0.01 ml/min at ambient temperature (~15 $^{\circ}$C). A 4× objective lens and 120keV (peak) were used for imaging. In total, 18 consecutive scans were obtained over a period of 46 hours. Each scan took around 2.5 hours and consisted of 1601 projections covering a rotation of 360°, with a voxel size of 3.2 μm. Prior to reconstruction, the interface of the aluminium wall of the core holder to the confining water was tracked and matched with a reference sinogram to compensate for drift and distortion caused by a changing drag force from the tubing during sample rotation. Ring artefacts were suppressed by Fourier-wavelet destriping of the sinograms[37] before reconstructing with the GridRec algorithm. After reconstruction, we applied 4 cycles of iterative non-local means denoising to every step of the time series, using a constant noise level estimate. The dataset was then aligned spatially, with digital volume correlation using Pearson's correlation coefficient as the quality metric. Details of the signal processing can be found in Bruns *et al*.[29, 38]

**Numerical Methods**

A model environment for the numerical simulation was created using X-ray holotomography reconstructions of drill cuttings from North Sea Hod chalk. We used a sample that was roughly cylindrical, with a diameter of ~500 μm. Data were collected at the ID22 beamline at the European Synchrotron Research Facility (ESRF) in Grenoble, France. We collected 1999 projections from a single dry scan at 29.5 keV, with an exposure of 500 ms and 360º rotation. The reconstruction had a voxel dimension of 100 nm.[39] For this study, the reconstructions were intensity aligned, compensated for ring artefacts, denoised with iterative nonlocal means algorithm and sharpened by deconvolution. An estimate of voxel level porosity was acquired by interpolating from an intensity based Gaussian mixture model.[29] An 11×11×26.4 μm³ domain containing ~3.2 million voxels was then cropped from the dataset and used for numerical study.

The laminar flow field is evaluated using Darcy's law. The volumetric flowrate between neighbouring voxels are given by

$$q = Rz \cdot \left(\frac{k(\varphi)}{l_n}\right) \cdot \Delta p, \quad (1)$$

where all the quantities are dimensionless. $q$ represents the flowrate, $l_n$ the voxel size, $\Delta p$ the pressure difference, $\varphi$ the geometric mean of the neighbouring voxels' porosity and $k(\varphi)$ the dependence of permeability on porosity. In this study we assume $k(\varphi) = \varphi^3 / (1-\varphi)^2 / \sum_1^6 |\Delta \varphi|_i$. The summation corresponds to the porosity differences between neighbouring voxels and $k$ is capped at $10^6$. $Rz$ is a dimensionless number that lumps the referencing quantities together

$$Rz = \frac{P_{ref} \kappa_{ref}}{L_{ref} v_{ref} \mu}, \quad (2)$$

where $P_{ref}$ represents pressure (Pa), $\kappa_{ref}$ permeability (mD), $L_{ref}$ length (m), $v_{ref}$ velocity (m/s) and $\mu$ viscosity (Pa·s). We solve the divergence-free continuity equation for each voxel

$$\sum_{i=1}^{6} q_i + q_s = 0, \quad (3)$$

where $i$ indexes the 6 neighbours of a voxel and $q_s$ represents the source term and is positive for voxels located at the fluid inlets and negative for those at the outlets. In this study we assign $q_s$ evenly over the inlet and outlet planes, equivalent to implementing a constant flow boundary condition in solving the Stokes equations. For a domain with $N=X\times Y\times Z$ voxels (where $X$, $Y$ and $Z$ represent the number of voxels along the Cartesian axes), $(4-1/X-1/Y-1/Z) \times N$ algebraic equations result from Equations 1 and 3 and are solved iteratively using the biconjugate gradient stabilized method.

The reactive transport of an aqueous species, A, is solved based on the flow field. The analytical form of the advection-diffusion equation is solved for fluxes across all voxel interfaces:

$$\left\{\exp\left[Pe \cdot \left(\frac{q \cdot l_n}{D(\varphi)}\right)\right] - 1\right\} \cdot n_A + q \cdot \left\{\exp\left[Pe \cdot \left(\frac{q \cdot l_n}{D(\varphi)}\right)\right] \cdot c_0 - c\right\} = q \cdot \left\{\exp\left[Pe \cdot \left(\frac{q \cdot l_n}{D(\varphi)}\right)\right] - 1\right\}, \quad (4)$$

where all quantities are dimensionless. $D(\varphi)$ represents the dependence of the effective diffusivity between neighbouring voxels on the geometric mean of their porosities. In this study we assume $D(\varphi)=\varphi^2$. $n_A$ represents the flux of A between neighbouring voxels, $c_0$ and $c$ are dimensionless concentrations of A in the two voxels. Pe is the Péclet number:

$$Pe = \frac{v_{ref} L_{ref}}{D_{A,ref}}, \quad (5)$$

where $D_{A,ref}$ represents a reference diffusivity of A. Pe is set to 1 for all simulations.

For the voxels where $\frac{q \cdot l_n}{D(\varphi)} < 10^{-6}$ we replace Equation 4 with

$$Pe \cdot \left(\frac{l_n}{D(\varphi)}\right) \cdot n_A + c_0 - c = 0 \quad (6)$$

to achieve better numerical stability. The voxels are modelled as continuously operating chemical reactors with perfect interior mixing:

$$\sum_{i=1}^{6} n_{A,i} + \left(Da \cdot k^I(c) \cdot \sum_{i=1}^{6} |\Delta\varphi|_i + q_{out}\right) \cdot c = -q_s + q_{in} \cdot c_s, \quad (7)$$

in which all quantities are dimensionless. The summations from 1 to 6 indicate quantities related to the reactor's neighbouring voxels. $q_{in}$ and $q_{out}$ represent the flowrates of sources and sinks, i.e. flows not coming from or going to any neighbouring voxel. $c_s$ represents the concentration of A in the source. The porosity difference, $\Delta\varphi$, appears because the surface area between neighbouring voxels is defined as $l^2|\Delta\varphi|$ (discussed in the next section). $k^I(c)$ represents the apparent first order rate constant defined by

$$k^I(c) = r_A(C_A) / k^I_{A,ref} / (C_{A,eq} - C_A), \quad (8)$$

where $C_A$ and $C_{A,eq}$ represent the concentration and the equilibrium concentration of A (mol/L), $k^I_{A,ref}$ the reference first order rate constant (m/s), $r_A(C_A)$ the reaction rate (mol/m$^2$/s) given concentration $C_A$. Reaction rate decreases as $C_A$ approaches $C_{A,eq}$ and $r_A(C_{A,eq}) = 0$. The dimensionless concentration is defined as $c = (C_{A,eq} - C_A)/C_{A,eq}$. Because $k^I$ depends, in general, nonlinearly on the concentration of A in each voxel, the $k^I$ and $c$ fields need to be solved iteratively.

Here, Da is the local Damköhler number

$$Da = \frac{k^I_{A,\text{ref}}}{v_{\text{ref}}} \quad (9)$$

and is different from the macroscopic Da in Table 1 which is discussed in the next section. For each species, Equations 4 and 7 yield a system of (4-1/X-1/Y-1/Z) × N algebraic equations. In this study, only the reactive transport of dissolved calcium (TOTCa) is simulated. The aqueous speciation in each reactor, including pH and the Saturation Index (SI) of calcite, are determined with PHREEQC using TOTCa as the master variable and assuming instantaneous aqueous equilibrium.[40] The derivation of the rate dependence on TOTCa, from which the function $k^I(c)$ is determined, is given in the SI (Equations S1-10, Figure S1). The parameters for the simulations are listed in Table 1. In the table, DI means deionized water of pH 7 is used as the flooding fluid. $CO_2$ means the solution is deionized water in equilibrium with 1 bar $CO_2$ and has an injection pH of 3.9. HCl means hydrochloric acid has been added to deionized water for pH adjustment so that it has the same injection pH as the $CO_2$ solution but differs in the apparent solubility for $CaCO_3$ ($TOTCa_{eq}$).

The porosity of each voxel is updated based on the pseudo steady state concentration field:

$$d\varphi = Ds \cdot \frac{k^I(c) \cdot c}{l_n} \sum_1^6 |\Delta \varphi|_i, \quad (10)$$

where $d\varphi$ is the porosity change over each time step. Ds represents the dimensionless time step

$$Ds = \frac{k^I_{A,\text{ref}} C_{A,eq}}{L_{\text{ref}} \rho_M} \cdot dt, \quad (11)$$

in which $\rho_M$ represents molar density (mol/L) and $dt$ the time step (s). The divergence free assumption in Equation 3 is valid when Ds is small. A constant Ds of $10^{-3}$ is used for all simulations in this study.

## RESULTS AND DISCUSSION

For the convenience of discussion, we define the Damköhler space (Da space) as the subvolume of a dissolving sample bounded between the pore wall and the dissolution front. In this study, both pore wall and dissolution front are isosurfaces defined by the chemical affinity of the dissolution reaction (instead of by the microstructure). Pore wall is the isosurface across which the chemical affinity of a dissolution reaction starts to decrease. Dissolution front is the isosurface across which the chemical affinity drops to 0 and the reaction reaches equilibrium. Damköhler space is the instantaneous reactive subvolume of a dissolving medium, i.e. 1) at any instant, only the geometric surface within the Da space is reactive; 2) any concentration variation of reactant or product outside the Da space stems from mixing.

Figure 1a to 1c show the evolution of pore morphology in 3D (top) and a cross section of the reconstructed microstructure normal to the flow direction (bottom). The 3D morphology is visualized by normalizing the voxel greyvalues of the complete time series and segmenting out the void by visual inspection (Figure S2). The same segmentation threshold is applied to all reconstructions. In the cross sections, the normalized greyvalues are presented without segmentation. The initial microstructure and the cross sections of the full time series are shown in Figures S3 and S4.

The shape of the developing pore is highly irregular and cannot be approximated by a simplified geometry with an analytical expression (e.g., a cylinder). We thus use the experimentally obtained greyscale tomograms in analysis whenever possible. The advancing speed of the pore tip – the foremost part of the pore space completely voided by chalk dissolution – was estimated to be 50 μm/h in the flow direction according to the segmented geometry. Beyond the tip, dramatic restructuring of pore space was observed within the Da space, where the preexisting, isolated small voids diminished and merged into a big, advancing pore. A rough estimate based on visual inspection suggested that the Da space was less than 10% of the cylindrical sample in all the timesteps, i.e. only a small portion of the field of view (FOV) was reacting at any given moment. The number of voids in the Da space decreased as the microstructure evolved, apparently contradicting the fact that the sample was dissolving. We attribute this contrary to the compaction of disintegrated structure and the relocation of very fine particles.[41] In addition, the disintegrating structure within the Da space can be difficult to visualize as the evolving geometric features got finer with time. However, the diminishing of old structure and the emergence of new geometric surface can be clearly identified by the spatial distribution of voxel greyvalue in each timestep.

Figure 1d and 1e show the evolution of local intensity variation, $\Delta I$(abs), along the flow direction. The summation sign on the $y$-axis indicates that the value is integrated over the plane at a given axial position. The differences between the plane and its neighboring, parallel planes are also accounted for. $I$(abs) represents the greyvalue of a voxel. It reflects the intensity of X-ray collected at the detector and corresponds to the X-ray absorption by – and therefore the electron density of – the sample. Quantifying $\Delta I$(abs) for neighboring voxels is essential for many tomography-based surface area (SA) measurement. For example, the value of $\Delta I$(abs) is directly related to the likelihood of two voxels being assigned to different phases if a dataset is segmented.[42] Consequently, most measurements based on binarised geometries, such as the advancing box approach, yield results correlative with $\Delta I$(abs).[43] In addition, methods based on two-point correlation probability function consider geometric surface the spatial variation of material density.[44] These methods yield results directly reflecting the frequency and amplitude of local intensity changes. A recent statistical analysis of petrophysical parameters of compacted carbonate based on greyscale tomography also used $\Delta I$(abs) as an important input in specific surface area (SSA) calculation.[29] In all cases, despite the use of various functional, the geometric SA show definite positive correlation with $\Delta I$(abs). In this study, we use $\Sigma|\Delta I$(abs)| as an indicator for SA without specifying a functional relationship. The result is qualitative and is subject to two technical issues that may lead to an underestimate of geometric surface in the FOV: 1) the surface momentarily generated and vanished within each timestep (~2.5 h) could not be captured; 2) many of the very fine and hierarchical heterogeneities in the natural chalk sample are beyond the characterisation resolution of our μCT. However, the trends in $\Delta I$(abs) evolution are clear and the time series suffices for a comparison with the numerical simulations.

The momentary increase of SA was observed to accompany the migration of the Da space in the flow direction. In Figure 1d, the horizontal dash line shows the average $\Sigma|\Delta I$(abs)| for an unreacted sample. The bold black line shows the sample profile in the first timestep. All curves fallen above this profile indicate the generation of SA during sample dissolution. The newly generated SA diminished with time, indicating that they contributed directly to the dissolution reaction and thus solid depletion, i.e. the new SA was *reactive*. In contrast to specific surface area, where a decrease in the normalizing denominator (e.g., sample mass) may lead to an increase of SSA despite a decreasing SA, $\Sigma|\Delta I$(abs)| correlates only with the SA and its increase cannot be attributed to the shrinking of grains.

The increase of SA is unlikely a result of reprecipitation. Throughout the experiment, the sample kept dissolving, evinced by a continuous and monotonic decrease of overall X-ray absorption and is clear from the visual inspection of the greyscale images (e.g., Figures S2 and 4). Before the arrival of the dissolution front, no significant morphological change was identified downstream, where the fluid was presumably saturated and precipitation would be favored. More importantly, precipitation decreases local permeability and redirects the flow to more porous regions.[7] This is opposite to what we observed in the time series, where the region with elevated SA was more reactive and drove the development of the advancing pore.

Local chemical heterogeneities may have contributed to the increase of SA. This increase may stem from the surface roughening accompanying the preferential removal of more reactive materials.[25, 41] For example, the biogenic calcium carbonate in chalk showed lower reactivity than the recrystallized calcite constituting the rest of the porous medium. These fossils hosted many of the originally isolated void spaces and may be bypassed by the developing pore (e.g., Figure 1d, the persistent peaks near 0.4 and 1.3 mm; corresponding fossilized structures visible in Figure 1a-c). However, SA increased through this mechanism is comparatively inert and would not migrate with the dissolution front, i.e. the SA should remain in the FOV and should not decrease over time. This anticipation is not consistent with our observation that most of the newly generated SA, while contributing to local structure disintegration, diminished with time.

We attribute the momentary increase of SA to the positive feedback between local permeability and local mineral dissolution rate.[45-47] This feedback constantly redirects less saturated fluid to more permeable flowpath. As a result, given a constant overall reactant influx, the more porous region of a medium gets more reactive fluid over time, further increasing its permeability and thus the chemical affinity of the passing fluid. The outcome of this feedback is the enhancement of local porosity contrasts. This contrast correlates positively to local density differences and therefore local SA. Because this positive feedback relies on the presence of a reaction whose rate depends on chemical affinity, the contrast enhancement may only occur within the Da space of a dissolving medium, i.e. all thus generated SA are reactive.

Figure 1e shows 5 selected timesteps during the microstructural evolution. The colored arrows indicate the relative increase of SA when compared to the previous timestep. The shaded areas show qualitatively the Damköhler length[48] (Da space in 1D) for the 3 intermediate timesteps, where the generation of reactive surface was significant. Similar increase of SA has been reported in an acidic system with elevated $CO_2$ pressure (~ 10 bars).[28] Our flooding fluid had an inlet pH of 7 and an effluent pH of 10. This circumneutral to alkaline condition suggests that the SA increase is not limited to the presence of acidic compound and may be applicable to a wide range of environmental scenarios where porous materials dissolve in flow field.[49] It is thus of interest to see how the SA increase can be affected by tuning the strength of the positive feedback through varying fluid composition or flowrate. Moreover, the positive feedback requires the initial presence of transport heterogeneity to demonstrate cumulative effect. Although heterogeneity is an intrinsic feature of natural porous media, tuning the feedback while keeping the initial heterogeneity constant is not experimentally feasible. Hence, the study of the surface generation process has to resort, at least partially, to numerical simulations based on a physically realistic microstructure.[24]

The trend of reactive surface evolution, recorded *in situ* during the percolation experiment, was also observed in all the numerical simulations in Table 1. Figure 2 shows a case study of the simulated percolation with ultrapure water (SimID 31222). We use the average porosity of the simulation domain ($\varphi$), instead of the elapsed time, as the measure of the overall reaction progress. This preference reflects the pivotal role of the macroscopic Damköhler number in determining the extent of surface generation and will be discussed with Figure 3.

The size of the digital chalk model used in the simulations is small (domain size: 11×11×26.4 $\mu m^3$, 3.2 million 32-bit greyscale voxels) compared to the experimental sample size. However, it captures the geometric features of a natural porous material with high fidelity. These complex features cannot be described analytically and are essential in studying the evolution of a realistic microstructure. The digital model allows us to study the positive feedback by using the exact and identical initial geometry while varying other operational parameters. Given that the intention of this study is not to propose a representative elementary volume (REV) for SA evolution, and that we do not study the effect of the scale dependent diffusive mixing (the Péclet number is set to 1 in all simulations), we consider the domain size sufficient in supporting our discussion.

In Table 1, the macroscopic Damköhler number is defined as

$$Da_M = k^I_{A,app} \cdot SSA \cdot \tau, \quad (12)$$

where *SSA* represents the volumetric specific surface area of the domain (m$^2$/m$^3$), and $\tau$ the mean residence time of the reactive fluid (s). The value of $Da_M$ can be kept constant by adjusting flowrate. The pseudo first order rate constant $k^I_{A,app}$ (m/s) is defined as

$$k^I_{A,app} = \left[ \int_0^{C_{A,eq}} \frac{dC_A}{r(C_A)} \right]^{-1}, \quad (13)$$

and reflects the amount of cumulative surface required by a solid-fluid reaction to reach the reaction front in a plug flow. Cumulative surface is the total availability of reactive surface to a flowing fluid parcel within the given residence time. The value of $k^I_{A,app}$ carries the information of the apparent solubility of a dissolving mineral in the flowing fluid, in the interval of integration, and shows the sensitivity of dissolution rate to the decrease of chemical affinity. For example, in Table 1, the dissolution of chalk in HCl solution is most sensitive to fluid composition change while it is least sensitive in the $CO_2$ solution. On a related note, the Damköhler number thus defined bears the physical significance of the ratio between the sample volume and the instantaneous volume of the Da space. For example, Da=20 means ~5% of the porous structure is dissolving.

The digital chalk model is considered chemically homogeneous, i.e. all solid materials within the simulation domain share the same dissolution rate law. This assumption rules out the possibility of surface roughening by preferential removal of more reactive substances.[41] The simulations also do not consider mineral precipitation or relocation of disintegrated grains. The microstructural evolution, therefore, stems only from solid dissolution in the imposed flow field. The flow field was evaluated by solving Darcy flow in the microstructure. We choose Darcy flow over solving Stokes flow in segmented microstructure because it allows 1) the access of fluid to the entire simulation domain and thus can simulate pore space perturbations (e.g., opening of new pores and/or breakthrough of reactive fluid) without *ad hoc* intervention with the constantly evolving flow field; 2) the incorporation of many more orders of magnitude of natural heterogeneities in assessing favorable flowpaths – with the same domain size, a 32-bit greyscale dataset can carry ~2 billion times more information than a binarised geometry. The second point is especially important for studying the instability of a migrating reaction front, where the presence of heterogeneity as perturbation can be essential.

The qualitative conclusions of the numerical simulation are not be affected by the selection of boundary condition or by the selection of the digital microstructure model. We imposed a constant volumetric flowrate on the domain so that the reactant influx did not change over time. We expect the results to be quantitatively different if a constant pressure gradient had been applied. The reactant influx shall increase as the domain becomes more porous, further accelerating structure disintegration. However, the choice of boundary condition does not affect the presence of infiltration instability, i.e. local increase of RSA will be observed with either selection of boundary condition. Similarly, a better characterization of the natural material (e.g., with a higher tomographic resolution or a greater FOV) will yield quantitatively more accurate simulation results. However, the validity of the coupled governing equations leading to an elevated local SA is not affected.

The evolution of pore morphology, in Figure 2a-c and in Movie S1, showed two important features along the flow direction. Downstream near the dissolution front and in the Da space, new, voided structure emerges from porous region of the sample, gradually turning into the tip of the advancing pore that drives fluid channelization. Meanwhile, upstream near the fluid entrance, newly formed flow channel expands its hydraulic dimension through the removal of materials at the solid-fluid interface. These two phenomena have sometimes been termed loosely as "wormholing" and the "expansion of existing conduit", and are considered fundamentally different.[50] However, in our studies they appear to coexist during porous structure development, but represent distinct localized morphologies.[24]

Figure 2d shows the evolution of mean SA per voxel in the axial direction. The SA is calculated from the porosity differences of neighboring voxels. For example, if the porosities of two neighboring voxels are 0 and 1, then this interface contributes $l^2$ to the overall SA, where $l$ is the voxel size (100 nm). The horizontal dashed line shows the average SA of the unreacted sample. The bold black line shows the profile of the unreacted sample and serves as a base line. The profiles fallen above this line all indicate the momentary increase of reactive surface. Given an axial position, the SA first increases because of the enhanced local porosity contrast, then decreases because the generated surfaces are reactive and accelerates solid depletion. This dynamic process is documented in Movie S2. Given any instant, this nonlinear behavior can only be observed within the subvolume of the simulation domain where the chemical affinity of the dissolution reaction changes. This subvolume, i.e. the Da space, moves in the flow direction and leads to an increase in the SA near the fluid outlet when $\varphi$=0.514 (Movie S3). The three vertical arrows in Figure 2e indicate the significant increases of SA in the three intermediate timesteps, as the Da space moves toward the exit.

The results of the 5 simulations in Figure 3 (Table 1) show the pivotal role of the macroscopic Damköhler number in determining the extent of SA increase. The conditions of the simulations were so chosen that 3 of them have the same inlet velocity while 3 have the same macroscopic Damköhler number, although the solution compositions differ. Figure 3a plots two petrophysical parameters, specific surface area (SSA) and overall porosity ($\varphi$), against elapsed time. The results differ significantly. The increase of porosity (solid lines) is determined by the dissolution capacity of the fluid flow, i.e. Flowrate×TOTCa$_{eq}$ in Table 1. As a result, the overall porosity increased fastest in SimID-67784 (CO$_2$, $Da_M$=2.2) and slowest in SimID-31222 (DI, $Da_M$=20). The SSA (dashed) showed momentary increase in all simulations. It is difficult to generalize the evolution of SSAs in 3a. In Figure 3b, however, the overall SA within the simulation domain is plotted against the dissolution progress. And it becomes clear that the extent of SA increase is determined by the volume of the Da space, i.e. given the same sample, the difference between its *in situ* SA and its *ex situ* SA is controlled by the Damköhler number defined in Equations 12-13. In the figure, SimID-67784 shows a distinct behavior than all the other simulations. This difference is because the $Da_M$ (2.2) is significantly lower in this simulation, meaning that the Da space consists of ~45% of the entire simulation domain. In contrast, the reactive subvolume is only 5% (for $Da_M$=20) and 4.3% (for $Da_M$=23) of the total volume in the other simulations. In the inset, the curves overlap with each other, with the one for SimID-53520 ($Da_M$=23) being slightly lower than the others because of the 0.7% smaller Da space.

This surprisingly good correlation between the SA evolution and the Damköhler number raised more questions than it has answered. It poses, paradoxically, the question that whether dimensionless numbers, such as Da or Pe, can really provide a quantitative framework to describe structural evolution on the pore scale.[11] Like most wormholing studies in the literature, the definition in Equations 12-13 1) does not reflect the heterogeneity of the medium and 2) depends on the sample size. One can safely assert that in a hypothetical medium that is perfectly homogeneous, the breakdown of reaction front symmetry will not be triggered with the same Da because no perturbation to the migrating front is present. As a result, no SA increase stemming from locally enhanced density contrast can be expected. Also, the residence time in Equation 12 scales with sample size. Thus defined Da is infinitely large for a domain with the same flow setup and cross section area, but infinitely long in the flow direction. However, at any instant, local structure changes occur only in the Da space, the volume of which is determined by the value of $k^I_{A,app}$. The porous medium beyond the dissolution front in the flow direction, while contributing directly to the calculation of Da, does not affect the instantaneous pore structure development. Hence, to claim a *quantitative* relation between pore structure development and Da, one has to always start with exactly the same domain (fixed size and identical internal geometry). This relation therefore has only limited practical value given the intrinsic heterogeneities of natural porous media. It is also interesting to notice that the physical significance of $Da_M$ – as the ratio of sample volume to reactive subvolume – is only valid if $k^I_{A,app}$ is determined using Equation 13. Any other choice of the rate constant will yield a different selection of flowrates for the same Da and will not produce the correlation in Figure 3b.

The existence of a SA($\varphi$)-Da correlation in greyscale simulations also implies a significant conceptual difficulty in understanding reactive infiltration instability from the perspective of a binary pore structure. This binary perspective means the internal geometry of a porous medium can always be fully resolved and sharp boundaries between solid and non-solid materials exist.[51] If so, the local porosity contrast is fixed: local porosity is either 0 or 1 and so is the porosity difference. Increase of SA in the Da space stemming from a local contrast enhancement is theoretically ruled out. As a result, the expansion of existing pores will dominate over the emergence of new pore parallel to the flow because streamlines do no penetrate through solid.

The increase of SA, if applicable, relies on specific geometric arrangements, such as the increase of hydraulic diameter of existing fluid conduits. An essential feature of the infiltration instability, that it amplifies any local heterogeneity through a positive feedback between flow and reaction, can be difficult to be reflected directly on the pore scale with a binarised structure.

The applicability of the SA($\varphi$)-Da correlation to various flow setups, e.g. conservative or divergent flow, remains to be investigated. Conservative flow means fluid flow through a conduit filled with porous media. The conduit has impermeable walls and a length much greater than its cross section dimension. The conservative flow is an idealization of core flooding or packed column experiments. If the Damköhler number is large and the flow is conservative, one would expect the amount of disappearing SA due to solid depletion be constantly compensated by the generation of SA as the Da space migrates continuously into unreacted region in the flow direction. Because the flow field is conservative, the Damköhler length does not change with the distance to the fluid inlet. The system thus may potentially have the remarkable feature that its *in situ* SA can reach a pseudo steady state that is **greater** than its *ex situ* SA. This nonlinear behavior is very different from the conventional understanding that the geometric SA (often measured *ex situ* using, for example, the B.E.T. method)[52] serves as the upper limit for reactive SA because of limited fluid accessibility.[18, 20]

The increase of SA may not be as significant if the flow is divergent. Divergent flow means a porous domain receives reactive fluid from a limited number of point sources and dissipates the reactants in a much larger space. This setup can be found in, e.g. geologic carbon storage in deep saline aquifers, where the injection wells serve as point sources and the acidic fluid, driven by pressure gradient, migrates away into the porous formation. The Damköhler number also approaches infinite in this case. However, as the distance from a point source increases, the flowrate decreases dramatically and the volume of the Da space gets rapidly compressed while the local Péclet number drops rapidly. As a result, local Da according to Equation 12 remains very large in a divergent flow whereas it can remain very small in a conservative flow. These new understandings may contribute to a change in the selection of the effective factor (a number that quantifies the difference between physical and reactive surface) that is often needed in environmental reactive transport modeling on large scales.[2]

The momentary increase of SA has 3 implications of environmental significance. First, the geologic self-organization induced by reactive infiltration instability is often associated with the presence of acidic environments.[10, 24, 25, 53, 54] Our results show that under circumneutral to alkaline conditions the instability can also be triggered. The different solutions used in our simulations and the consistency of the results suggest that the key parameter of the evolutionary process is not the rate of mineral dissolution far from equilibrium, but how a reaction approaches equilibrium[55] and as a result how large the Da space can span. Our conclusion may be relevant for basic environments such as those involved in nuclear waste disposal, where the hydroxyl-mediated dissolution of silicate is significant.[34, 35] Second, while visualizing a reactive aqueous species is difficult, contemporary tomography techniques can routinely capture geometric features of solid materials.[14] The correlation between SA and the Da space provides a new strategy to visualize reaction front in natural porous materials without introducing tracers that may complicate aqueous speciation. Third, during *in situ* remediation of migrating contaminant plume, it is desirable to know a treatment's region of influence.[56] A momentary increase of SA implies a temporary shrinking of the influenced region because the reactivity of a remediating fluid depletes faster with greater reactive surface area. This temporary migration of reaction front against the hydraulic pressure gradient may further complicate field engineering.

**FIGURE CAPTIONS**

**Figure 1**. Pore structure evolution in a cylindrical chalk sample (ø=0.9 mm, L=2 mm) percolated by ultrapure water under ambient condition. (a)-(c) 3D morphology of segmented pore structure (top) and a cross section of reconstructed greyscale tomogram (bottom) at different elapsed time. (d) Axial profiles of integrated local intensity gradient indicating the evolution of internal surface area. (e) Profiles of 5 selected time steps. Shaded regions show qualitatively the instantaneous Damköhler length in which the increase of surface area is indicated by vertical arrows.

**Figure 2**. Microstructure evolution in a simulated ultrapure water percolation (SimID 31222, domain size 11×11×26.4 µm$^3$). (a)-(c) 3D morphology of segmented pore structure (top, threshold porosity = 0.5) and a cross section of the microstructure (bottom) at different overall porosity. Overall porosity is used as a measure of reaction progress. The greyvalue corresponds to voxel porosity. The dataset was segmented for 3D structure visualisation only. All simulations were based on a greyscale tomogram. (d) Axial profiles of surface area at different reaction progress. (e) Five selected profiles with vertical arrows indicating momentary increase of surface area. See also Movies S1-S3.

**Figure 3**. Results of the simulations in Table 1. All simulations used the same initial microstructure. (a) Evolution of volume averaged specific surface area and overall porosity with time. (b) Evolution of surface area with overall porosity, which is used as a measure of dissolution progress. Colour coding is shared in both figures.

**TABLES**

**Table 1.** Parameters used in the numerical simulations.

| SimID | Solution | pH$_0$ | pH$_{eq}$ | $k_{A,app}^{I}$ [m/s] | TOTCa$_{eq}$ [mol/m$^3$] | Flowrate [m/s] | Da$_M$ |
|---|---|---|---|---|---|---|---|
| 31222 | DI | 7.0 | 9.8 | 2.78E-06 | 0.122 | 4.98E-06 | 20 |
| 90831 | HCl | 3.9 | 9.4 | 3.32E-06 | 0.185 | 5.76E-06 | 20 |
| 68168 | CO$_2$ | 3.9 | 6.0 | 3.15E-07 | 9.31 | 5.46E-07 | 20 |
| 53520 | HCl | 3.9 | 9.4 | 3.32E-06 | 0.185 | 4.98E-06 | 23 |
| 67784 | CO$_2$ | 3.9 | 6.0 | 3.15E-07 | 9.31 | 4.98E-06 | 2.2 |

# FIGURES

Figure 1

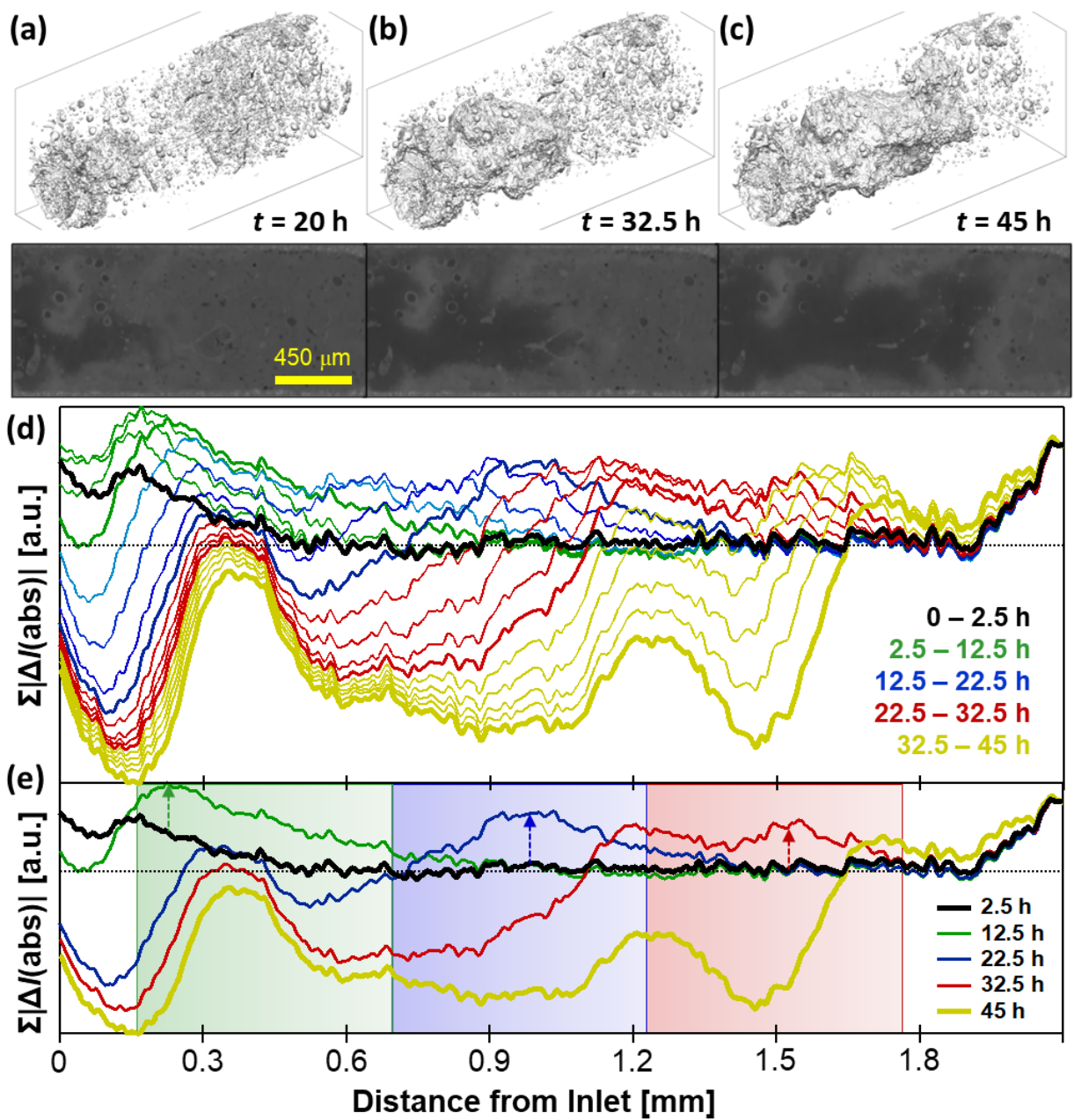

Figure 2

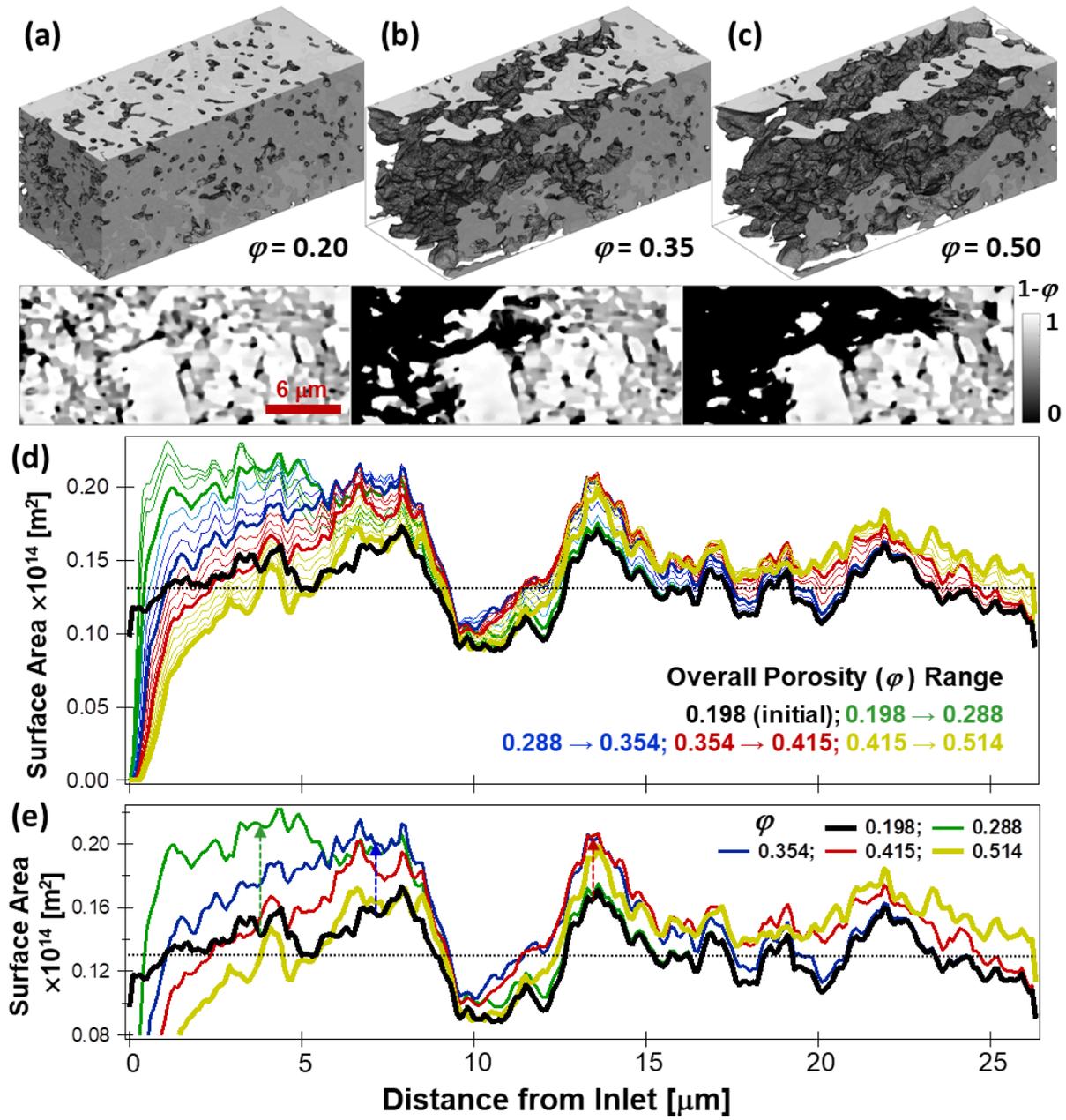

Figure 3

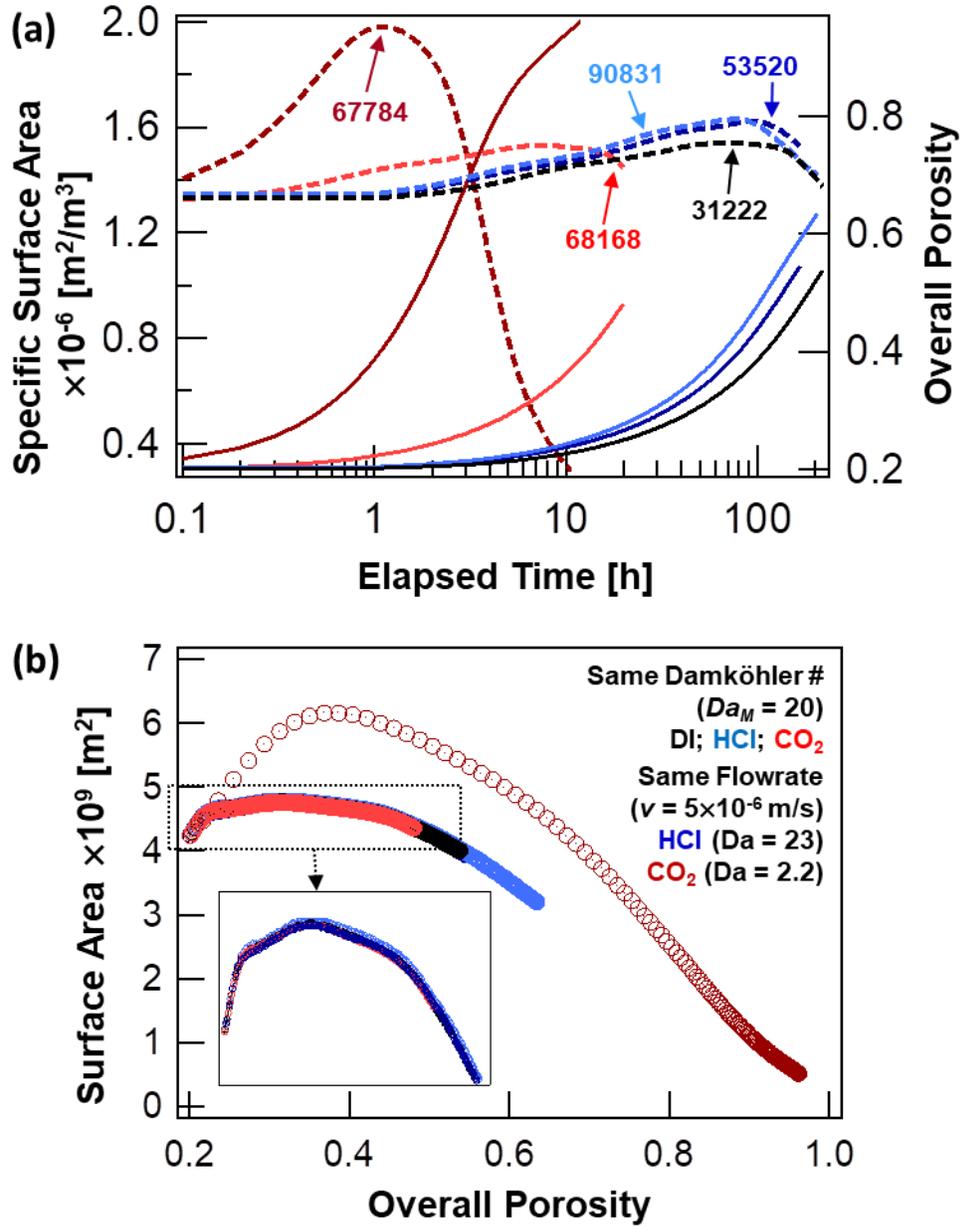

## ASSOCIATED CONTENTS

**Supporting Information.** Rate dependence on TOTCa. 3D visualisation of segmented pore structure for all *in situ* μCT time steps. The initial microstructure. A cross section time series showing pore development over the 45 hour experiment with an FOV of 1x2 mm$^2$.

**Supplementary Movies.** Three supplementary movies (Movies S1-S3) are deposited in ERDA, the Electronic Research Data Archive maintained by the University of Copenhagen, and are accessible through link https://sid.erda.dk/wsgi-bin/ls.py?share_id=DCF6f256oY

**Movie S1.** Microstructure evolution. SimID 31222. Perspective view. Domain size 11×11×26.3 μm$^3$.

**Movie S2.** Evolution of reactive surface. SimID 31222. Cross section along the flow direction. Field of view 11×26.3 μm$^2$.

**Movie S3.** Evolution of Damköhler space. SimID 31222. Perspective view. Domain size 11×11×26.3 μm$^3$.

## ACKNOWLEDGMENTS

We thank F. Saxild for help with design and manufacture of the percolation cell. Funding for this project was provided by the European Union's Horizon 2020 Research and Innovation Programme under the Marie Sklodowska-Curie Grant No 653241, the Innovation Fund Denmark through the CINEMA project and the Innovation Fund Denmark and Maersk Oil and Gas A/S through the P$^3$ project. Parts of this work were also supported by funding from the European Commission for the CO2-REACT project, Grant Agreement No: FP7-317235. We gratefully acknowledge support for synchrotron beamtime from the Danish Agency for Science, Technology and Innovation via DANSCATT.

Supporting Information for

# Momentary increase of reactive surface in dissolving carbonate

Y. Yang,[*,1] M. Rogowska,[1] Y. Zheng,[2] S. Bruns,[1] C. Gundlach,[2] S. L. S. Stipp[1] and H. O. Sørensen[1]

[1]Nano-Science Center, Department of Chemistry, University of Copenhagen, DK-2100 Copenhagen, Denmark

[2]Department of Physics, Technical University of Denmark, DK-2500 Lyngby, Denmark

[*]Corresponding author, +45 9161 4575, yiyang@nano.ku.dk

9 Pages, 4 Figures

## Rate Dependence on TOTCa

The dependence of chalk dissolution rate on the total aqueous concentration of Ca (TOTCa) was measured by dissolving 1.2 g of chalk powder (grainsize ~500 μm) in 600 mL continuously stirred ultrapure water in a batch reactor. The pH evolution was monitored continuously and was used to determine the evolution of TOTCa with PHREEQC. Three initial fluid compositions were used, corresponding to the 3 scenarios in Table 1 (DI, HCl and $CO_2$). In each experiment, the solution was sampled 10 times during the first hour of dissolution and the TOTCa was measured using an atomic absorption spectrosopy (PerkinElmer Aanalyst 800, Waltham, MA, USA) to validate the PHREEQC calculation. At any given instance, TOTCa in a batch reactor is governed by

$$V \frac{dC_A}{dt} = \frac{N_A}{\tau_A}, \quad (S1)$$

where the subscript A refers to total calcium, $V$ represents the volume of the solution (m³), $t$ the dissolution time (s), $N_A$ the number of reactive sites on the mineral surface (mol) and $\tau_A$ the characteristic time (s) of calcium release. $\tau_A$ is proportional to the reciprocal of the intrinsic reaction rate and depends only on the composition of the solution. $N_A$ is time dependent and reflects the dynamics between the removal of reactive sites by dissolution and the emergence of new sites by surface renewal. Hence, we look for a solution in the form $N_A = N_{A,ss} + N_A(t)$, where $N_{A,ss}$ represents a steady state solution and $N_A(t)$ is a transient term that describes the reduction of available reactive sites during dissolution. Let $N_{A,ss} = \chi_\infty N_{A0}$, where $N_{A0}$ is $N_A$ at $t=0$ and $\chi_\infty$ the percentage of renewable $N_A$ as $t \to \infty$. The governing equation for $N_A$ is therefore

$$\frac{dN_A}{dt} = -\frac{N_A(t)}{\tau_A}, \quad (S2)$$

which gives

$$N_A = \chi_\infty N_{A0} + (1-\chi_\infty)N_{A0} \cdot e^{-\int_0^t \frac{dt}{\tau_A}}, \quad (S3)$$

leading to

$$\tau_A \left( \frac{dC_A}{dt} \right) = \frac{\chi_\infty N_{A0}}{V} + \frac{(1-\chi_\infty)N_{A0}}{V} \cdot e^{-\int_0^t \frac{dt}{\tau_A}}. \quad (S4)$$

This integral differential equation regarding $\tau_A$ is a Fredholm equation of the second type. It can be solved by discretising the exponential term on the right hand side

$$\exp\left(-\int_0^{t_1} \frac{dt}{\tau_A(t)}\right) \exp\left(-\int_{t_1}^{t_2} \frac{dt}{\tau_A(t)}\right) \cdots \exp\left(-\int_{t_{n-1}}^{t} \frac{dt}{\tau_A(t)}\right) \approx \left(1-\frac{\Delta t_1}{\tau_{A,1}}\right)\left(1-\frac{\Delta t_2}{\tau_{A,2}}\right) \cdots \left(1-\frac{\Delta t_n}{\tau_{A,n}}\right), \quad (S5)$$

where *n* is the total number of data points. Back substitution gives a system of *n* quadratic equations that can be solved recursively

$$\left.\frac{dC_A}{dt}\right|_{t_1} \cdot \tau_A^2(t_1) - \frac{N_{A0}}{V}\tau_A(t_1) + (1-\chi_\infty)\frac{N_{A0}}{V}\Delta t_1 = 0$$

$$\left.\frac{dC_A}{dt}\right|_{t_2} \cdot \tau_A^2(t_2) - \left[\chi_\infty\frac{N_{A0}}{V} + (1-\chi_\infty)\frac{N_{A0}}{V}\left(1-\frac{\Delta t_1}{\tau_A(t_1)}\right)\right]\cdot\tau_A(t_2) + (1-\chi_\infty)\frac{N_{A0}}{V}\left(1-\frac{\Delta t_1}{\tau_A(t_1)}\right)\Delta t_2 = 0$$

...

$$\left.\frac{dC_A}{dt}\right|_{t_i} \cdot \tau_A^2(t_i) - \left[\chi_\infty\frac{N_{A0}}{V} + (1-\chi_\infty)\frac{N_{A0}}{V}\cdot\prod_{k=1}^{i-1}\left(1-\frac{\Delta t_k}{\tau_A(t_k)}\right)\right]\cdot\tau_A(t_i) + (1-\chi_\infty)\frac{N_{A0}}{V}\cdot\prod_{k=1}^{i-1}\left(1-\frac{\Delta t_k}{\tau_A(t_k)}\right)\cdot\Delta t_i = 0$$

(S6)

and

$$\tau_A(t \to 0) = \frac{H + \sqrt{H^2 - 4\cdot(1-\chi_\infty)H\cdot\Delta t_1\cdot\left(\left.\frac{dC_A}{dt}\right|_{t=0}\right)}}{2\cdot\left(\left.\frac{dC_A}{dt}\right|_{t=0}\right)}, \quad (S7)$$

where $H = SLR \cdot SSA \cdot \sigma_A$ is determined experimentally. *SLR* represents the solid to liquid ratio (g/m$^3$), *SSA* the specific surface area (m$^2$/g) and $\sigma_A$ the density of surface sites (mol/m$^2$). $\Delta t$ is the time interval between 2 consecutive pH recording and is 1 second in this study.

Similarly,

$$\tau_A(t_i) = \frac{\chi_\infty H + H_i + \sqrt{(\chi_\infty H + H_i)^2 - 4\cdot H_i\cdot\Delta t\cdot\left(\left.\frac{dC_A}{dt}\right|_{t_i}\right)}}{2\cdot\left(\left.\frac{dC_A}{dt}\right|_{t_i}\right)}, \quad (S8)$$

where

$$H_i = (1-\chi_\infty)H\cdot\prod_{k=1}^{i-1}\left(1-\frac{\Delta t_k}{\tau_A(t_k)}\right). \quad (S9)$$

Once $\tau_A(t_i)$ and $N_A = V(\chi_\infty H + H_i)$ are obtained by monitoring the concentration variation $\left(\left.\frac{dC_A}{dt}\right|_{t_i}\right)$ in a batch reactor, the chalk dissolution rate can be written as a function of total dissolved calcium ($C_A$) by noticing the monotonic mapping between $C_A$ and the elapsed time (*t*)

$$C_A(t) = \lim_{\Delta t \to 0} \left\{ \int_0^t \frac{SLR \cdot SSA \cdot \sigma_A}{\tau_A(t)} \left[ \chi_\infty + (1-\chi_\infty) \cdot \prod_{k=1}^{i-1} \left(1 - \frac{\Delta t_k}{\tau_A(t_k)}\right) \right] \cdot d\dot{t} \right\}, \qquad (S10)$$

where $\dot{t}$ is a dummy variable.

The evolution of dissolution rate and TOTCa is plotted in Figure S1. The rate dependence on TOTCa is determined from Figure S1b and c.

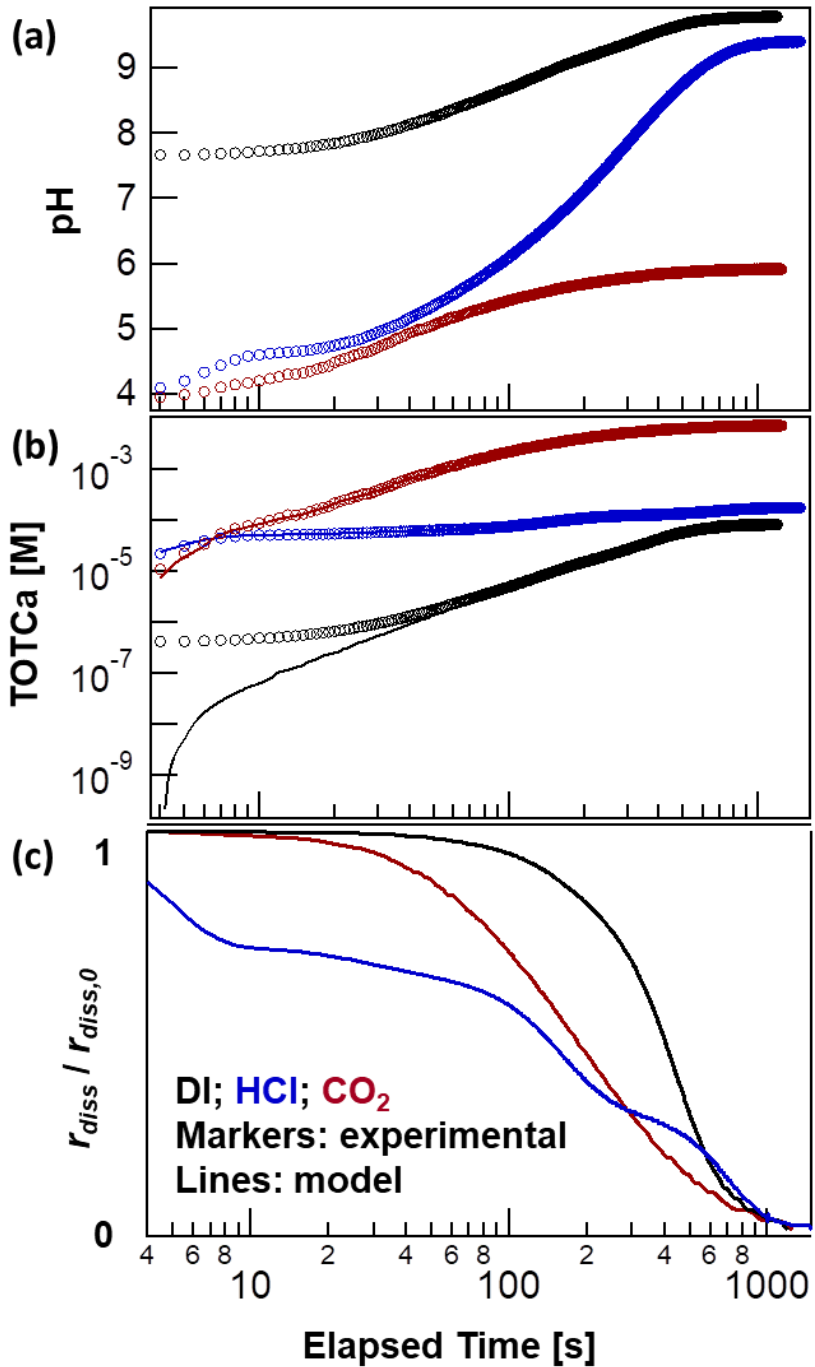

**Figure S1**. Results of closed free drift dissolution experiments which are used to determine rate dependence on total dissolved Calcium (TOTCa). (a) Evolution of system pH recorded in the experiment. (b) Evolution of TOTCa. Markers are values determined with PHREEQC from measured pH. Solid lines are calculated using Equation S10. (c) Evolution of dissolution rate determined from Equations S7-9 as the reciprocal of the characteristic time $\tau_A$.

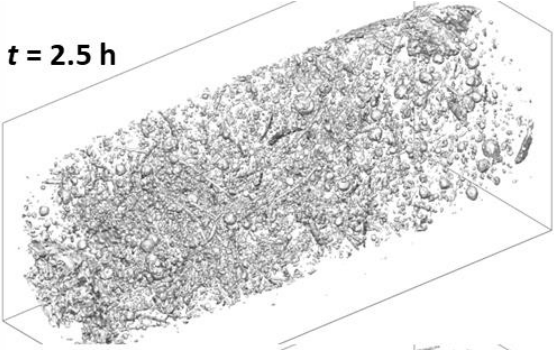
*t* = 2.5 h

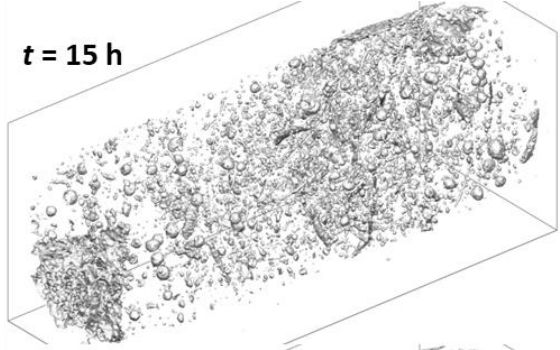
*t* = 15 h

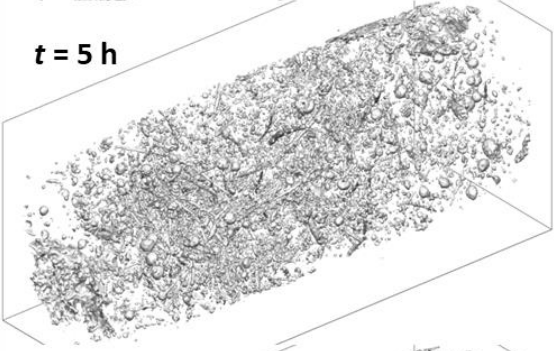
*t* = 5 h

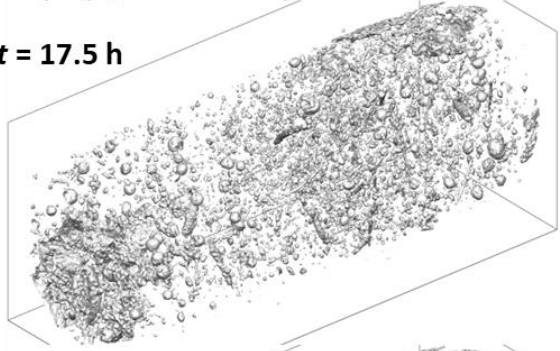
*t* = 17.5 h

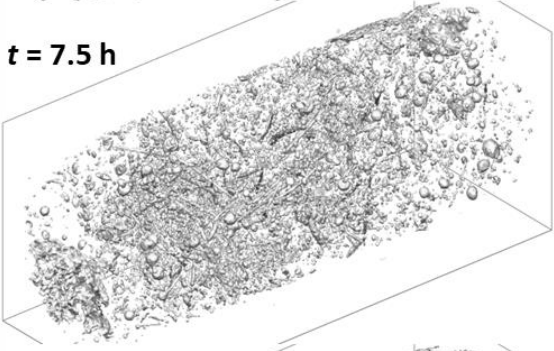
*t* = 7.5 h

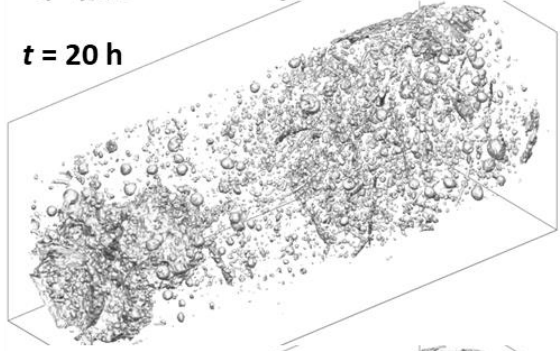
*t* = 20 h

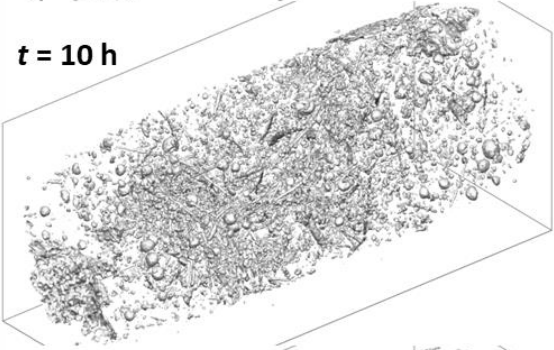
*t* = 10 h

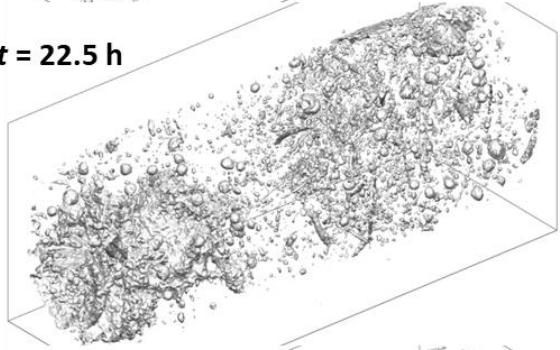
*t* = 22.5 h

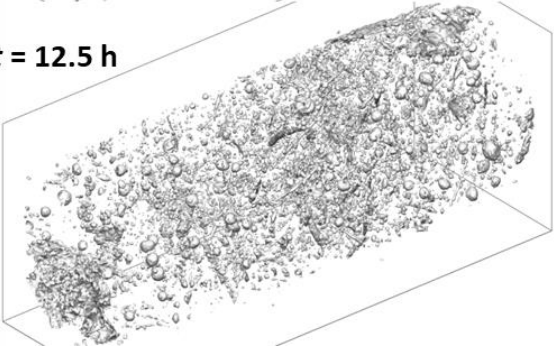
*t* = 12.5 h

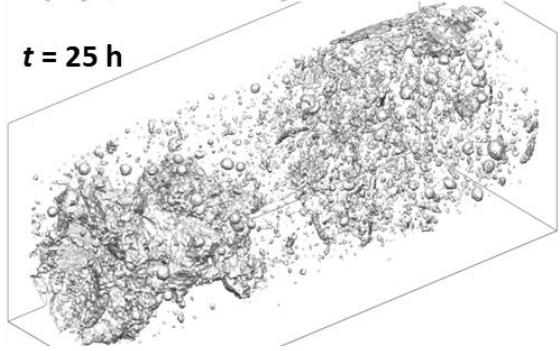
*t* = 25 h

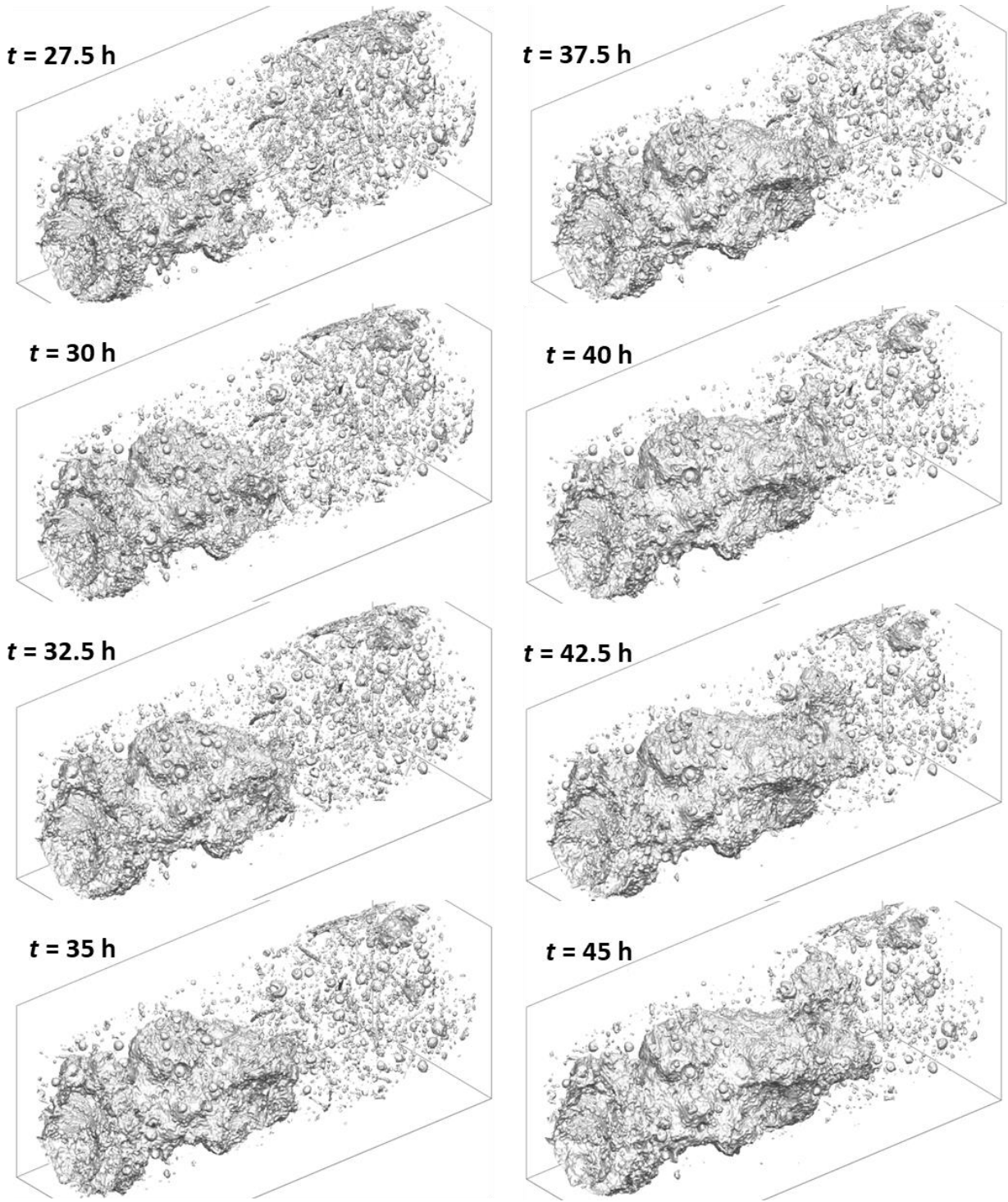

**Figure S2**. Evolution of the segmented pore structure from the *in situ* μCT recording.

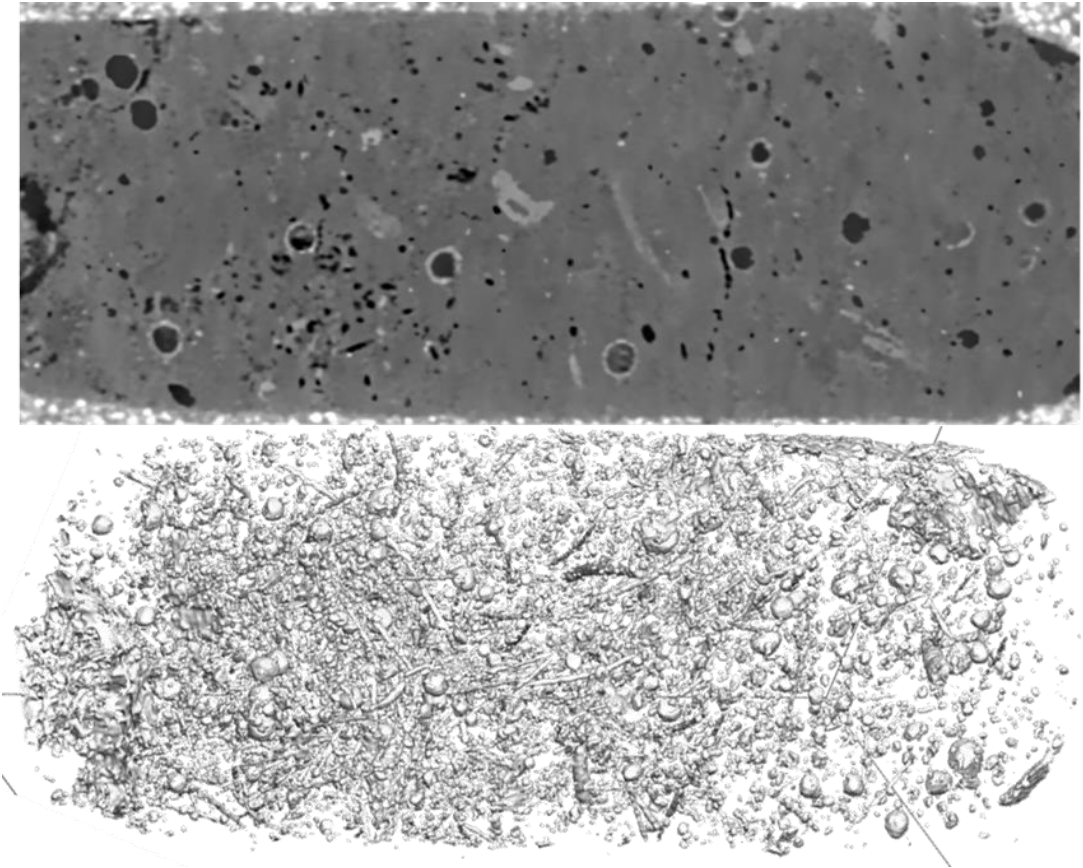

**Figure S3**. Initial microstructure of the chalk sample used in the *in situ* measurement. Top: a cross section in the axial direction; Bottom: segmented pore structure (pre-existing voids).

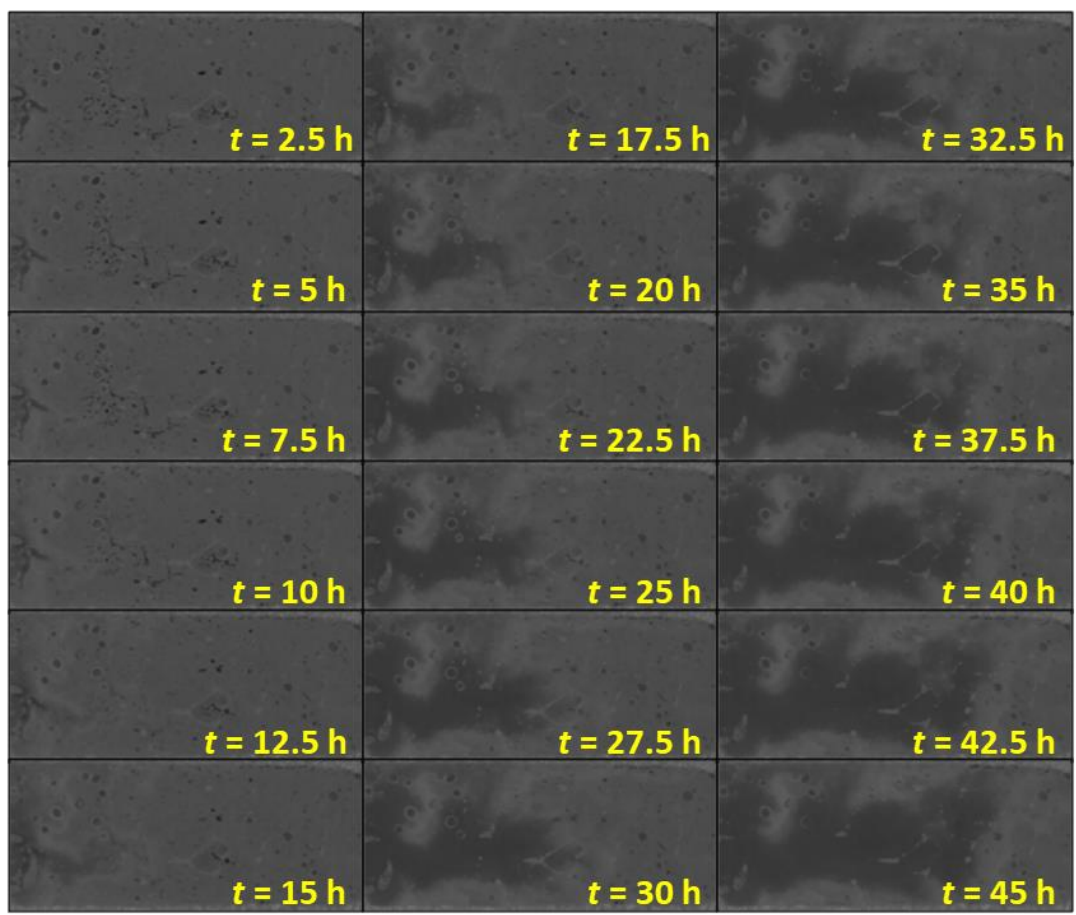

**Figure S4.** A time series of greyscale cross section showing the evolution of pore structure over ~45 hours. Each image corresponds to an FOV of 1x2 mm$^2$.